# On Lee Smolin's
# *The Trouble with Physics*

**Jean-Paul Auffray**
Ex CIMS-NYU
jpauffray@yahoo.fr
(july 2007)

**Abstract**. Lee Smolin's casual accounting of special and general relativity in *The Trouble with Physics* raises an interesting question: is it possible to develop a legitimate argument concerning string theories starting from a shaky basis? This is apparently what Lee Smolin succeeded in doing when he wrote *The Trouble with Physics*. The book's shortcomings are nevertheless troublesome.

## INTRODUCTION

I wish to cast a light on Lee Smolin's recent publication, *The Trouble with Physics* [1]. Lee Smolin expresses in this book his disappointment with the current status of string theories. The book contains a remarkable 18-pages Index which refers both to names and to concepts alluded to in the book. This Index proved to be a valuable tool for probing the book's contents on specific questions. I report here some preliminary observations thus obtained..

## 1. SMOLIN ON SPECIAL RELATIVITY

Not counting Notes and the Index, *The Trouble with Physics* covers about three hundred and fifty pages. References to special relativity in the Index point to forty-three pages in the book. A reading of the forty-three pages shows that Smolin's understanding of special relativity is strongly Einstein-oriented. This is true both in regard to the history of the theory and to its physical content.

    Einstein is indeed a major figure in *The Trouble with Physics*. His name occurs at least once on nearly one hundred pages of the book. It is unclear why. Einstein was never involved in anything resembling strings. His long-standing rejection of relativistic spacetime and his life-long lack of acceptance of the quantum theory as it developed during his lifetime are well known.

    Given the interest for Einstein's life and work demonstrated in *The Trouble with Physics*, Lee Smolin might have been well advised to inform his readers of the way special relativity came into being in 1905.

# 2. ON THE BIRTH OF SPECIAL RELATIVITY

Then working as an Examiner third class at the Swiss Federal Bureau of Intellectual Property, Albert Einstein sent his first paper on relativity to *Annalen der Physics* on June 30, 1905. In this paper, he expounds a line of reasoning which allegedly leads to the coordinate transformation equations we know today as the Lorentz transformation [2]. Einstein destroyed his manuscript shortly after his paper appeared in print. And he subsequently abandoned the line of reasoning he had proposed in this paper to establish the Lorentz transformation. No major physics textbook – not even Abraham Pais's mythical *Subtle is the Lord...*[3] – has ever taken the pain to reproduce Einstein's original line of reasoning. Einstein himself never returned to it – for reasons that Lee Smolin might have wished to explain to his readers.

# 3. SMOLIN ON GENERAL RELATIVITY

With regard to special relativity, Lee Smolin has of course a great excuse. He simply reiterates in his book the specious views physicists at large hold concerning the way relativity came into being. Special relativity was established in full by Henri Poincaré before Einstein. It is a sad reality few physicists, if any, are willing to contemplate [4].

Less understandable is Lee Smolin's presentation of general relativity in *The Trouble with Physics*. His book Index shows a single reference to David Hilbert whose work on general relativity preceded Einstein's in 1915.

When Einstein decided to enter the physicist's fray to construct a relativistic theory of gravitation, he asked his good friend Marcel Grossmann to identify for him the mathematical tools that would be required to bring the project to fruition. A brilliant mathematician in his own rights, Grossmann rose to the occasion and in no time constructed the desired theory – but Einstein rejected it on the basis of an "intuitive" consideration concerning the principle of "Unicity" [5].

Einstein's departure for Berlin and the eruption of World War I in 1914 separated the two friends.

In June of 1915, Einstein paid a visit to the mathematics department at the University of Göttingen. There, he met David Hilbert, who had become the leading figure in the world of mathematics after Henri Poincaré's untimely death in 1912. Too old to serve in the military, Hilbert had remained behind even as many of his best students enlisted for service in the German armed forces.

Hilbert invited Einstein to stay at his house while he was in Göttingen. He listened patiently as Einstein explained to him his fundamental "discovery" that a generally-covariant theory of gravitation was a mathematical impos-sibility.

After Einstein left Göttingen to return to Berlin, Hilbert went on a trip to the Baltic island of Rügen. Returning to Göttingen on November 14, 1915, he sent a letter to Einstein, telling him with great pride and excitement: "I have found an axiomatic solution to your great problem!" And he sent to him a copy of the solution he had found.

Einstein lacked Hilbert's fabulous mathematical skills, but he was more "clever" than him. He manoeuvred skilfully to make it appear that Hilbert, not Einstein, had "plagiarized" the other.

Einstein's biographers are embarrassed when it comes to describe faithfully what actually happened that November between Göttingen and Berlin. Yet, it is all very clear. As Albrecht Fölsing explains it in his recently published comprehensive Einstein's biography: "In November [1915], when Einstein was totally absorbed in his theory of gravitation, he essentially corresponded only with Hilbert […]. On November 18, [he thanked him] for a draft of his treatise." Then comes the pertinent question: "Could Einstein, writes Fölsing, casting his eye over Hilbert's paper, have discovered the term which was still lacking in his own equation and thus 'appropriate' Hilbert?" [6]

Of course not, suggests Fölsing. It just happened that the missing term found its way quite naturally into Einstein incomplete equations. This is what genius is all about!

Hilbert presented his relativistic theory of gravitation under the name *Grundlagen der Physik* in Göttingen, on November 16, 1915. Einstein followed suit in Berlin with his own theory – which contained at last the "missing term" – nine days later.

Lee Smolin writes on page 45 of his book: "In 1915, Einstein had written to David Hilbert […]: 'I have often tortured my mind in order to bridge the gap between gravitation and electromagnetism.'."

In formulating his relativistic gravitation theory, Hilbert had in fact gone further than Einstein was able to go in Berlin nine days after him in two fundamental respects: he had founded his construction on the discovery of a principle of least action which applies to the whole universe and he had included the electromagnetic field in his considerations – two decisive aspects of the theory which are entirely missing in Einstein's version of it.

## 4. GENERAL RELATIVITY'S THIRD MAN

At the university La Sapienza in Rome in 1975, professor Reno Ruffini sought to correct the injustice which has made Marcel Grossmann the forgotten "third man" of general relativity. To that effect, he created an international Committee, which has organized an international "Marcel Grossmann Meeting" every three years ever since and attributed prizes to selected institutions or persons who successfully advanced the cause of general relativity. The prize bears Grossmann's name not Einstein's.

There is not entry for Marcel Grossmann in Lee Smolin's Index – nor, for that matter, for Felix Klein or for Emmy Noether who, together with David Hilbert, set general relativity straight [7].

## 5. SMOLIN ON ACTION

Another great absent in Lee Smolin's otherwise fascinating book to read is the concept of action. There is no entry for action between "acceleration" and "Adelberger, Eric" in Lee Smolin's Index. And yet, action is at the very heart of string theories. It might be said that string theories are in fact quantum theories of action.

Lee Smolin's apparent lack interest in action has led him to make odd statements in his book such as this: "[…] Nature produces collisions between particles and antiparticles. They annihilate, creating a photon." [8]. Physics textbooks usually present this type of collision in terms such as these: "*Two photons* [my emphasis] of equal energy, moving in opposite directions, are produced."

If Lee Smolin's description of a one-photon annihilation process corresponds to observed events, then a deep revolution is imminent in particle physics.

## 6. REALITY OR A MIRROR EFFECT?

Is Lee Smolin's view of what contemporary physics is all about precisely as his pen makes it appear throughout his book? Or did his pen betray him on occasions, causing him to express his thoughts not quite as he had whished to do initially…?

In *The Trouble with Physics*, Lee Smolin's pen sometimes appears indeed to have plaid tricks on him. This would seem to be the case, for example, in this presentation of the inner workings of special relativity: "According to it [special relativity], the geometry of space is that given by Euclid […]; however, space is mixed with time, in order to accommodate Einstein's two postulates." [9]

In elementary school in my days we were told that one cannot mix oranges with bananas. And yet physicists apparently have discovered that space "is mixed" with time.

Lee Smolin's pen also appears to have led him astray when he wrote this odd presentation of general relativity on page 81 of *The Trouble with Physics*: "Einstein revealed that the geometry of space is evolving in time, according to other, deeper laws." [10] If the geometry of space can evolve "in time", then time exists independently of space and we are back to Newton's physics.

# Conclusion

In spite of its obvious shortcomings, *The Trouble with Physics* is a book worth reading as it expounds for the benefit of a large public, with undisguised candour, the hopes, aspirations and frustrations young contemporary physicists experience when they contemplate their apparent inability to match the great achievements physicists of the previous generations were able to produce in their time when they forged the quantum theory and relativity.

Lee Smolin's casual accounting of special and general relativity in *The Trouble with Physics* raises an interesting question: is it possible to develop a legitimate argument in physics today starting from a shaky basis?

This is apparently what Lee Smolin succeeded in doing when he wrote *Te Trouble with Physics*.

I am less pessimistic than Lee Smolin when it comes to evaluate the results obtained so far by his generation: string theories have not yet cast their ultimate shadow.

If one is permitted to conclude these candid remarks on a humorous note, I will dare say that Smolin's perusal account in *The Trouble with Physics* of Einstein's contributions to special and general relativity illustrates the ever-lasting pertinence of a famous remark Richard Feynman made during a lecture he delivered in Los Angeles in 1981 when he said: "By the way, what I have just outlined is what I call 'a physicist's history of physics,' which is never correct. [It is] a sort of conventionalized myth-story that the physicists tell to their students, and those students tell to their students, and is not necessarily related to the actual historical development, which I do not really know." [11]

*The Trouble with Physics*'s most interesting parts occur early in the book when Lee Smolin expounds what are, according to him, the twenty-first century "Five Great Problems in Theoretical Physics". I intend to present a study of the five problems in a forthcoming publication [12].